\documentclass[final, 3p]{elsarticle}

\usepackage{lineno}
\usepackage[colorlinks=true,breaklinks=true,pdftex]{hyperref}
\modulolinenumbers[100]
\usepackage{booktabs}
\usepackage[T1]{fontenc}
\usepackage{pgfplots}
\pgfplotsset{height=5cm, compat=1.8}
\usepgfplotslibrary{statistics}
\usepackage{mathtools}
\usepackage{hyperref,amsmath,amsfonts,amssymb,amsthm,soul, subfigure}
\usepackage{graphicx,epstopdf,epsfig,color,float,caption}

\theoremstyle{remark}
\newtheorem*{remark}{Remark}

\journal{GMP 2021}

\biboptions{authoryear}

\graphicspath{{./figures/}}

\def\RR{\mathbb{R}}
\def\ZZ{\mathbb{Z}}
\def\YYY{\mathcal{Y}}
\def\ZZZ{\mathcal{Z}}

\begin{document}
\begin{frontmatter}

\title{Binary segmentation of medical images\\ using implicit spline representations and deep learning}

\author[SINTEF]{Oliver J.D. Barrowclough}
\ead{oliver.barrowclough@sintef.no}

\author[SINTEF]{Georg Muntingh\corref{cor}}
\ead{georg.muntingh@sintef.no}

\address[SINTEF]{Department of Mathematics and Cybernetics, SINTEF, Forskningsveien 1, 0373 Oslo, Norway}

\author[IVS]{Varatharajan Nainamalai}
\ead{rajan.riks@gmail.com}
\address[IVS]{The Intervention Centre, Rikshospitalet, Oslo University Hospital, 0372, Oslo,  Norway} 

\author[SINTEF]{Ivar Stangeby}
\ead{istangeby@gmail.com}

\cortext[cor]{Corresponding author}

\begin{abstract}
We propose a novel approach to image segmentation based on combining implicit spline representations with deep convolutional neural networks. This is done by predicting the control points of a bivariate spline function whose zero-set represents the segmentation boundary. We adapt several existing neural network architectures and design novel loss functions that are tailored towards providing implicit spline curve approximations.
The method is evaluated on a congenital heart disease computed tomography medical imaging dataset. Experiments are carried out by measuring performance in various standard metrics for different networks and loss functions. We determine that splines of bidegree $(1,1)$ with $128\times128$ coefficient resolution performed optimally for $512\times 512$ resolution CT images.
For our best network, we achieve an average volumetric test Dice score of close to 92\%, which reaches the state of the art for this congenital heart disease dataset.
\end{abstract}


\end{frontmatter}

\section{Introduction}

Image segmentation is the process of partitioning an image of pixels into different regions according to shared attributes. It is a technology widely used in multiple fields ranging from self-driving vehicles and surveillance to medical imaging; see \cite{minaee2020} for a recent comprehensive survey on image segmentation using deep learning.

Image segmentation techniques can be divided into three types: manual, semi-automatic, and fully automatic \cite{ISIN2016317}. In manual segmentation, trained people classify the region of interest one image at a time. An automatic image segmentation followed by a manual threshold or a manual seeding of a region growing algorithm is referred to as semi-automatic image segmentation. Fully automatic segmentation, mainly based on machine learning (ML) methods, applies prior knowledge based on learned features to new unseen images. 

Manual segmentation of images is a time-consuming and tedious job, so there is potentially great value in developing methods that can partially or fully automate the process. Semi-automatic and automatic methods for image segmentation date back several decades and include approaches such as thresholding, region growing, classifiers, clustering, Markov random-field models, deformable models, atlas-guided approaches and (artificial) neural networks \cite{Pham2000}. In recent years, there has been an explosion in the use of neural networks, especially in the field of computer vision, where they typically significantly outperform other approaches.  

One common feature of image segmentation is that the segmentation boundaries are often smooth, while at the same time exhibiting complex topological behaviour. This is particularly true in the case of medical imagery, but is also typical in many other types of image data. Complex topologies can occur, for example, when slicing a 3D object with a simpler topology, or from objects being partially occluded in photographs. Spline representations are well known in the field of computer-aided design for providing excellent compact representations of smooth geometries. Implicit representations, which represent a curve or a surface by the zero-set of a multivariate function, are well suited for modelling complex topologies and for visualization by ray tracing. By using tensor-product splines as an implicit function, we combine the respective benefits and obtain a representation that compactly models a wide variety of segmentation boundaries.

Splines have previously been used in the context of deep learning, but as far as we are aware, not for implicitly representing segmentation boundaries. For example, Catmull-Rom splines are employed by \cite{ling2019fast} to define segmentation boundaries as parametric curves. 
Deep learning has also been used to compute parametrizations that can be used for parametric spline approximation \cite{laube2018deep}. One weakness of the parametric approach compared to the implicit approach is that it is difficult to model complex topologies with parametric geometries. Splines have also been used in a slightly different context, to extend convolutional neural networks (CNNs) (see Section \ref{sec:CNN}) to irregular and geometric datasets \cite{fey2018splinecnn} by defining continuous B-spline kernels. Similarly, more general continuous spline representations of the weight space have been considered \cite{keskin2018splinenets}, using splines to compactly model entire filter banks and fully connected layers.

The use of implicit representations in deep learning has appeared in a number of recent works involving a variety of problems. 
The problem of recovering a 3D model from a single image of an object using implicit representations has been investigated by several authors \cite{NIPS2019_8340, Michalkiewicz2019}, achieving state-of-the-art performance on certain datasets.
Implicit representations based on neural networks with periodic activation functions have also been considered for modelling images, videos and sounds along with their derivatives \cite{sitzmann2020implicit}, addressing issues with the level of detail provided by conventional implementations. 
Other works using levels sets of spline functions for the purpose of image segmentation have appeared in the context of evolutionary processes \cite{yang2006evolution}.

Within the field of medical imaging, segmentation is used for studying anatomical structures, estimating the volume of certain tissues, planning treatments and post-treatment follow-ups, and identifying and monitoring the development of tumors, lesions and abnormalities \cite{neeraj2010}.
There exist a number of non-invasive imaging techniques that can be subject to segmentation, including ultrasound, magnetic resonance imaging (MRI) and computed tomography (CT) imaging.
In this paper, we focus on the problem of segmenting CT images from a dataset of congenital heart disease (CHD) patients \cite{miccai2019}. A CHD is a structural birth defect in the heart, or blood vessels near the heart, that can disrupt the normal flow of blood.

In this paper we combine implicit representations and deep convolutional neural networks into a new method for image segmentation. The method is applicable to general image segmentation problems, and we show that it reaches state-of-the-art results on a CHD CT image dataset. The novelties of this paper include:
\begin{itemize}
    \item A new end-to-end procedure, based on deep convolutional neural networks, for segmenting images with implicitly represented splines that compactly represent smooth and topologically complex segmentation boundaries.
    \item Several neural network architectures based on existing networks, including truncated VGG-style networks \cite{simonyan2014very} and an adaptive version of UNet \cite{ronneberger2015u}. This new network is adaptive both in the sense of being able to take variable size input for fixed output (via repeated application of adaptive average pooling) and in the sense that it can adapt to different gridded basis functions (in our case, tensor-product splines).
    \item New loss functions that are tailored to the specific problem of modelling implicit splines, by mapping control points to a binary inside-outside mask.
    \item A parameter study that determines the uniform tensor-product spline spaces with optimal performance for the CHD CT image dataset.
\end{itemize}

Our implementation of the networks, loss functions, and training and data processing procedures is based on PyTorch. This code is open source and available as a GitHub repository \cite{Barrowclough2020}.

\section{Background}
\subsection{Deep learning based whole heart segmentation}
There have been some attempts to obtain ML-based automatic CT and MRI cardiac image segmentation for full blood volume or for heart chambers. \cite{miccai2019} used a UNet-based deep learning network for CHD CT images of 68 volumes and obtained an average Dice score of 0.7843 and 0.773 for blood volume and myocardium, respectively (2D UNet for blood volume segmentation, 3D UNet for chambers and
myocardium segmentation). \cite{Rajan} used a 3D DenseVNet CNN model from \cite{Gibdvnet} for the same CHD CT image dataset, and obtained a mean Dice score of 0.9183 and 0.8519 for blood volume and myocardium, respectively. A few authors have used the cardiac CT angiography dataset from the MICCAI 2017 Multi-Modality Whole Heart Segmentation Challenge, consisting of seven structures of heart, including the left ventricle, myocardium of the left ventricle, left atrium, right ventricle, right atrium, pulmonary artery and the ascending aorta. \cite{payer} used a 3D UNet architecture model with bounding box around all heart structures, and obtained a mean Dice score of 0.889 for the whole heart CT image segmentation. \cite{Wang} used a 2D UNet with shape context estimation. \cite{CFUN} used a faster R-CNN and 3D UNet networks for the whole heart CT image segmentation. \cite{Habijan} used a 3D UNet architecture CNN model with principal component analysis as a data augmentation technique, and obtained an average Dice score of 0.89 for the whole heart CT image segmentation.

\subsection{Convolutional neural networks}\label{sec:CNN}
In this section, we recall the basics of CNNs, necessary for describing the truncated VGG-style \cite{simonyan2014very} and UNet-style \cite{ronneberger2015u} architectures used in this paper.

Given an image $I$ and a \emph{kernel} (or \emph{filter}) $K$, each taking real values on a finite subset of $\RR^2$, their \emph{(discrete) convolution} is defined as a new image
\begin{equation}\label{eq:convolution}
O = I*K := \left(\sum_m \sum_n K(m,n) I(i-m, j-n)\right)_{i,j}.
\end{equation}
Flipping the minus signs one obtains a \emph{cross-correlation}. In our setting, the kernel typically has small support, and the cross-correlation measures, in each location in the image, to what extent the image locally correlates with the kernel. A simple example is the kernel $K$ whose only nonzero values are $K(0,0) = -1$ and $K(1, 0) = 1$, compactly denoted by $K=[-1, 1]$, measuring the forward difference of the image in the first coordinate. The resulting output image $O$ then measures, at each pixel, the presence of a vertical edge (called a \emph{feature}) in the input image. The values of $K$ are often called \emph{weights}.

Many neural network libraries implement cross-correlations but call them convolutions, and this is the core ingredient of a CNN. A \emph{convolutional layer} typically takes many convolutions (with different filters) of the same input image in parallel, with resulting output images stacked as \emph{channels} of a triple array, each measuring the presence of a different feature in the input image. The input image can also consist of multiple channels, as is the case for instance for RGB images. In our case, each filter is a triple array, with two spatial dimensions and one channel dimension. Confusingly, it is common practice to omit the channel dimension in describing the size of the filter. For instance, a kernel of size $1\times 1$ actually has a single entry for every channel, and it is used to take linear combinations between image channel values in corresponding spatial locations. In particular in the case of RGB images, the kernel $K = [1,0,0] \in \RR^{1,1,3}$ would measure the presence of the ``redness feature'' in each pixel.

In theory, it is possible to consider a single convolutional layer (\emph{shallow learning}) with a vast number of filters that together account for any conceivable pattern in the image, but in practice this is not very efficient. The success of CNNs is to instead consider hierarchies of features that build complex features out of simple features (\emph{deep learning}). For instance, if our first convolutional layer measures the presence of a horizontal line $-$ and a vertical line $|$ in separate channels, then in a subsequent layer the $1\times 1$ kernel $K=[1,1]\in \RR^{1,1,2}$ is able to measure the presence of a $+$. Being linear maps, simply composing filters just yields another filter, typically with fewer degrees of freedom than the sum of those of the individual filters. Instead, features of higher complexity are obtained by adding a \emph{non-linearity} (an \emph{activation function}) in-between subsequent layers, such as the Rectified Linear Unit (ReLU, defined as \textbf{1}$_{>0} \cdot x$) or the hyperbolic tangent ($\tanh$). 

Instead of hand-crafting these filters (called \emph{feature engineering}), deep learning obtains them by minimizing a loss function expressing the discrepancy between the network prediction and a given label (the \emph{ground truth}). The loss function is often minimized using a stochastic variant of gradient descent, where the gradient of the loss for the entire dataset is estimated by the loss for a representative subset (\emph{mini-batch}). Such noisy gradients are desirable for escaping local minima and, more importantly, saddle points, which are prevalent in high-dimensional weight space. This gradient is then distributed across the weights using the chain-rule, in a process called \emph{back-propagation}. When `unfolding' the gradient in this manner, earlier layers can sometimes receive too little (\emph{vanishing gradient}) or too much (\emph{exploding gradients}). This is resolved by adding a \emph{batch normalization} layer in between convolutional layers, which shifts the distribution of a batch of inputs to each layer by applying a normalizing filter. Another technique for making the gradient propagate more effectively through the network are \emph{skip connections}, which connect early layers (almost) directly to deeper layers (cf. Figure~\ref{fig:architectures}, bottom).

While deep hierarchical representations dramatically reduce the number filters required for solving many tasks, still an enormous number of activation function values (\emph{activations}) need to be stored in memory for learning these filters, especially when images are processed in parallel in (mini-)batches. In order to make deep networks feasible on existing hardware, the spatial resolution can be reduced. This is further justified by effectively increasing the \emph{receptive field} of deeper features, referring to the pixels in the input image that are involved in its computation. In addition, it makes the features invariant to local perturbations, which is often considered a positive side effect.

One method for downscaling the spatial resolution is \emph{pooling} spatial groups of features. The standard example involves dividing the image spatially into blocks, and replacing each block by the maximum (\emph{max pooling}) or average (\emph{average pooling}) of its feature values. This can be done either directly by specifying the block size, or \emph{adaptively} by specifying the desired output resolution. Alternatively, the spatial resolution can be reduced using \emph{strided convolutions}, which compute \eqref{eq:convolution} only at pixels $(i,j)$ in some lattice in $\ZZ^2$, for instance $2\ZZ \times 2\ZZ$.

Repeatedly down-scaling is useful in object detection, where one desires a single scalar output interpreted as the probability of the presence of an object in the input image. However, for segmentation, an output resolution comparable to the input resolution is required. Observe that all convolutions are linear maps, with sparsity pattern determined by the support of the kernel and stride. Hence the adjoint of a strided convolution (i.e., the linear map with transposed matrix) maps to a higher dimensional space corresponding to a larger image. These learnable up-scalings are called \emph{up-convolutions} (or \emph{transposed convolutions} or \emph{convolutions with fractional stride}). The UNet incorporates these convolutional layers, first in a \emph{contracting path} with $d$ down-scalings (the \emph{depth}) resulting in a layer with the (\emph{bottleneck}) $b\times b$ spatial resolution, followed by an \emph{expanding path} involving $d$ up-scalings. More detail is provided in Section \ref{sec:networkarchitecture}.

\begin{figure}[ht]
    \centering
    \includegraphics[scale=0.7]{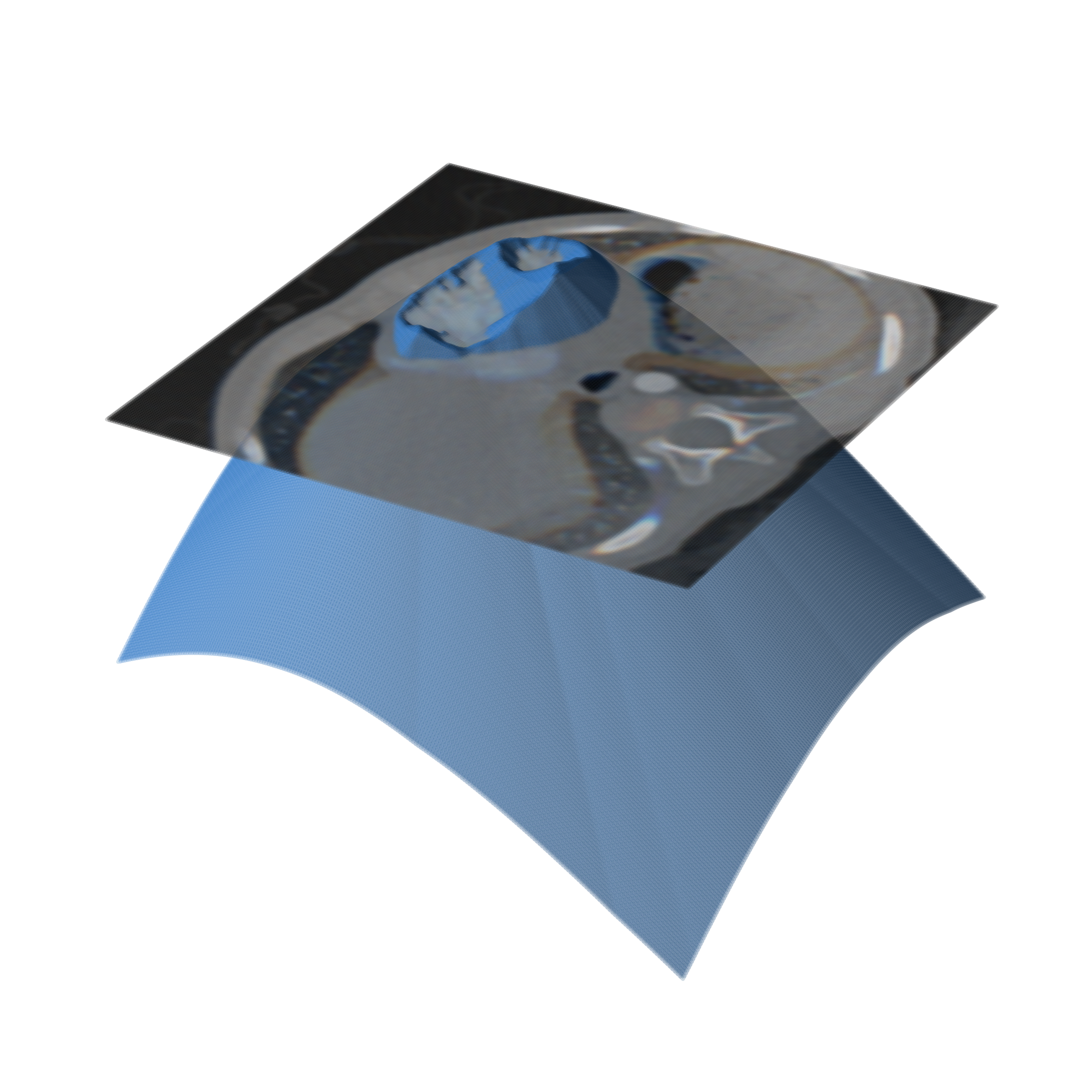}
    \caption{A CT slice segmentation represented as an implicit function, here shown with outside as negative for visualization purposes.}
    \label{fig:implicitrep}
\end{figure}

\subsection{Geometric representations for segmentation}\label{sec:geometricrepresentations}
Previously, segmentation data has been discretely represented as boundary polygons \cite{PolyRNN} or as graphs \cite{acuna2018efficient}. Binary segmentation masks, or full segmentation maps with resolution corresponding to the input image have also been considered \cite{ronneberger2015u}. In the situation that the underlying topology is known, active contouring has also been used for boundary segmentation \cite{aubert2003image}.

Alternatively, a \emph{smooth} geometric representation for segmentation boundaries of \emph{unknown topology} can be provided by implicit functions~$F$. Here the boundary is not modeled using an explicit parametrization, but implicitly as the points $(x,y)$ in the plane for which $F(x,y)=0$. In addition, the points $(x,y)$ inside the segmentation satisfy $F(x,y)<0$, while those outside the segmentation satisfy $F(x,y)>0$; sometimes this convention is reversed, see Figure \ref{fig:implicitrep}.

The function $F$ can be modelled in many ways, including with multivariate polynomials, radial basis functions, (signed) distance fields and with splines. In this paper, we choose such implicit functions from appropriate tensor product spline spaces, which, due to their gridded nature, are well suited for representing smooth geometries implicitly as the output of fully convolutional neural networks. The implicit spline representation has the following additional features:
\begin{itemize}
    \item \emph{Representation}: It is compact, capable of representing complex topology, has a built-in degree-dependent continuity establishing a smoothness prior on the segmentation. It can also represent features of arbitrarily small size, and is thus not limited by pixel resolution. 
    \item \emph{Processing}: Inside/outside computations are reduced to mere function evaluations, allowing for instance for swift volume computation; manipulation of the shape and comparison between shapes is efficient in terms of the spline control net; derivatives, tangent vectors and normal vectors are readily computed from the implicit form, yielding offsets and confidence bounds for the modelled shapes.
\end{itemize}

\subsection{Spline modelling}\label{ssec:splinemodelling}
Bivariate tensor-product splines are widely used for smooth surface approximation, due to their high approximation order and intuitive control net-based representation. For a given degree $p$ and number of basis functions $O$ (to be used for the \emph{output} representation), consider a non-decreasing sequence $(t_i)_{i=1}^{O+p+1}$, whose entries are called \emph{knots}. The constant B-spline basis functions $B_{1,0}, \ldots, B_{O,0}$ are defined by
\[
B_{i,0}(x) := \left\{
\begin{array}{ll}
1     &  t_i \leq x < t_{i+1},\\
0     & \text{otherwise}.
\end{array}
\right.
\]
The higher-degree B-spline basis functions $B_{i,p}(x)$ are defined recursively as
\[ \frac{x - t_i}{t_{i+p} - t_i} B_{i,p-1}(x) + \frac{t_{i+p+1} - x}{t_{i+p+1} - t_{i+1}} B_{i+1, p-1}(x), \]
with the convention that $0/0$ evaluates to $0$.

For a fixed degree $p$ and (open) knot vector
\[
t_1 = \cdots = t_{p+1} < t_{p+2} < \cdots < t_O < t_{O+1} = \cdots = t_{O+p+1},
\]
the set
\[ \mathcal{B} := \{B_{i,p}(x)\,:\,1\leq i\leq O\}\]
forms a basis of the vector space of $C^{p-1}$-smooth functions on the interval $[t_1,t_{O+p+1}]$ restricting to polynomials of degree $p$ on the intervals $[t_i, t_{i+1}]$.

Using the tensor-product construction, the above univariate B-spline basis gives rise to a bivariate B-spline basis
\[ \mathcal{B} \otimes \mathcal{B} := \{ B_{i,p_1}(x) B_{j,p_2}(y)\,:\,1\leq i,j\leq O\} \]
of the vector space of $C^{p_1-1,p_2-1}$-smooth functions restricting to polynomials of bidegree $(p_1,p_2)$ on the rectangles $[t_i, t_{i+1}]\times [t_j, t_{j+1}]$.

In this paper we consider several different bidegrees, but due to the nature of the data, we restrict our attention to symmetric bidegrees of the form $(p,p)$. We also consider different resolutions $O\times O$ of output spline coefficients, focusing on the cases $O=64,128$ and using \emph{open uniform} knot vectors
\[
(t_i)_{i=1}^{O+p+1} = 
[\underbrace{0,\ldots,0}_{p+1\text{ times}},1,2,\ldots,O-p-1,\underbrace{O-p, \ldots, O-p}_{p+1\text{ times}}].\]
Any $C^{p-1,p-1}$-smooth piecewise polynomial of bidegree $(p,p)$ on this knot vector takes the form
\begin{equation}\label{eq:BivariateSpline}
F(x, y) = \sum_{i=1}^O \sum_{j=1}^O c_{i,j} B_{i,p}(x) B_{j,p}(y),
\end{equation}
for certain \emph{spline coefficients} $c_{i,j}$ arranged in a rectangular grid of control points, also known as a \emph{control net}. Our implementation works with batches of such coefficient grids, arranged as a triple array $C = (c_{b,i,j})_{b=1,i=1,j=1}^{B,O,O}$ for a batch of size $B$.

\section{Methodology}\label{sec:methodology}
In this paper, we propose a method for image segmentation, by combining implicit spline representations with fully convolutional neural networks. Due to the gridded nature of the coefficients, tensor-product splines are well suited for representing smooth geometries implicitly as the output of fully convolutional neural networks.

\subsection{Data preparation pipeline}
The medical image CT dataset we consider consists of 66 volumes that capture images of and around patients' hearts \cite{miccai2019}. The data has resolution $512 \times 512$ with a varying number of slices from 130--340 in the $z$-direction. The pixel spacing is 0.25 mm $\times$ 0.25 mm in each slice and 0.5 mm between slices. Manually labelled segmentation maps are available for each volume for two tissue categories: blood volume and myocardium.
We divide the segmentation maps into binary masks, one set for each of the categories.
Based on \cite{Rajan}, we have split the 66 volume dataset into 13 volumes for validation (volumes 3, 6, 8, 10, 17, 31, 32, 33, 45, 50, 52, 54, 68), 14 volumes for testing (volumes 2, 12, 18, 23, 25, 27, 40, 44, 48, 51, 53, 55, 57, 60), and the remaining 39 volumes (except for volumes 43 and 62, which have different resolution and slightly erroneous labels) for training.
Each volume is converted to a set of two-dimensional slices, each of resolution $512 \times 512$. This is done both for the image and mask volumes.

Our initial experiments involved an extra step, where we computed new spline coefficient ground truths as approximations of the original mask data. To do this, we performed a weighted least-squares approximation on a signed distance field computed directly from the mask using a fast marching method. However, this approach made experimentation cumbersome, as each change in spline resolution or degree required a computationally heavy recompute of distance fields and spline approximations. Eventually, we abandoned this approach in favour of one that includes spline evaluation as part of the loss function. In this way the masks provide the ground truth, despite the network outputting spline coefficients of a lower resolution. This approach will be described in more detail in Section \ref{ssec:loss}.

\begin{figure*}[t!]
    \includegraphics[height=4.6cm]{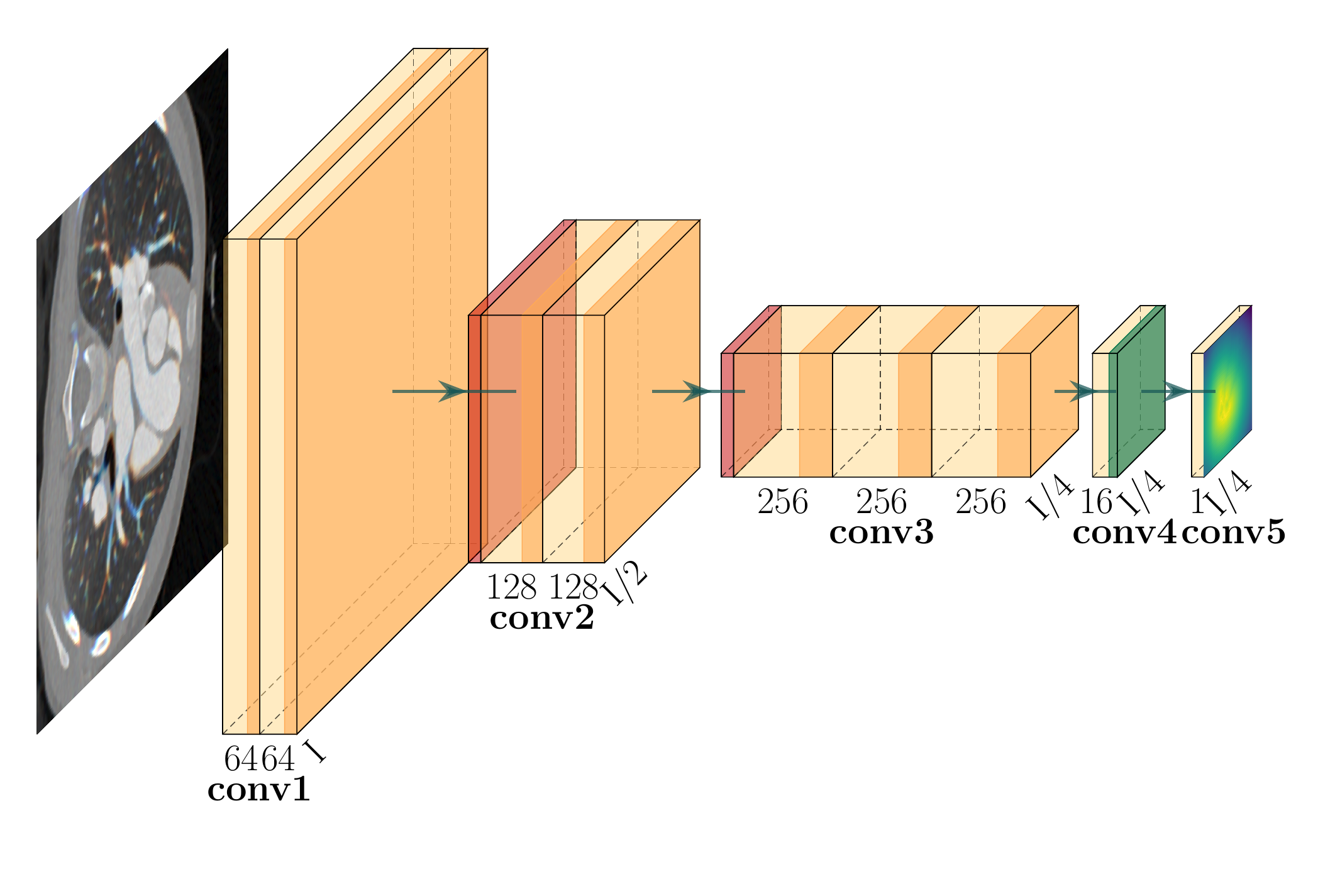}
    \includegraphics[height=4.6cm]{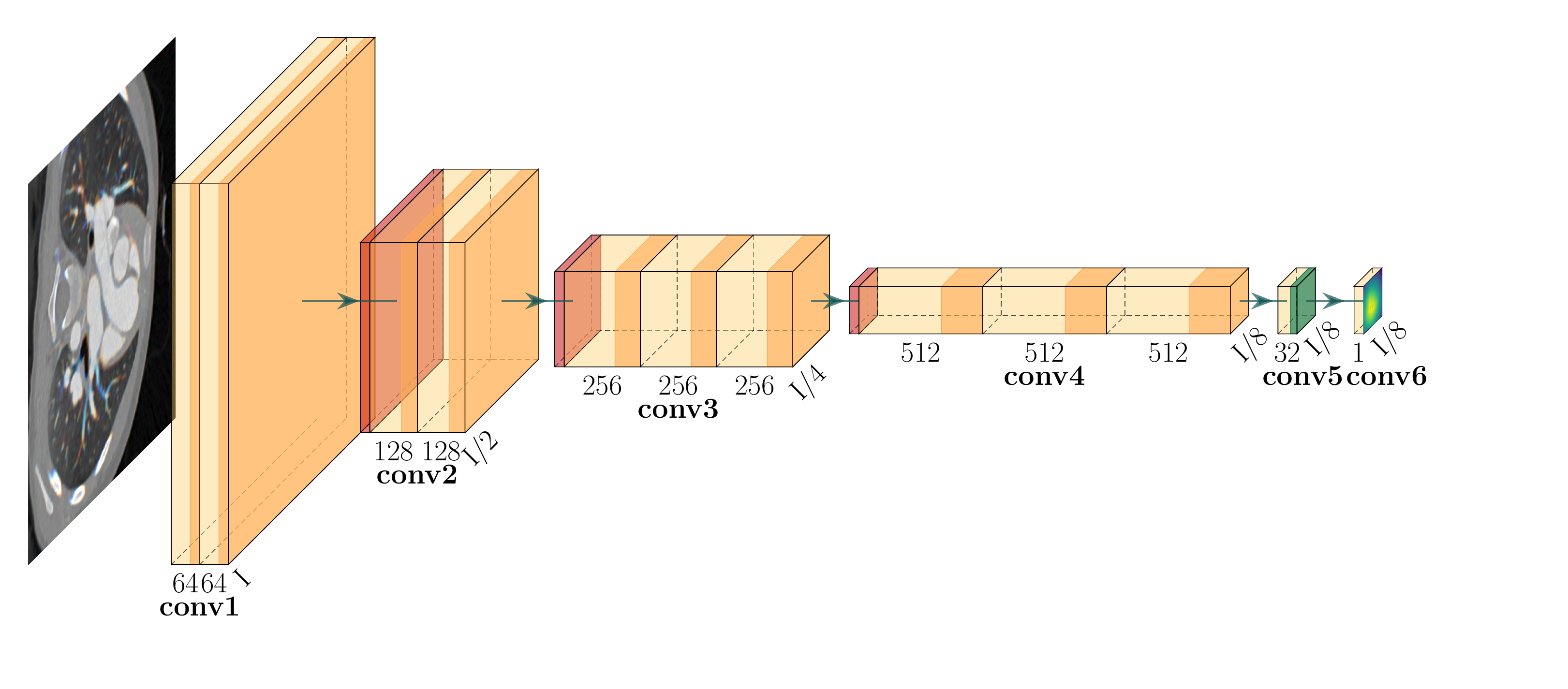}
    \includegraphics[height=4.6cm]{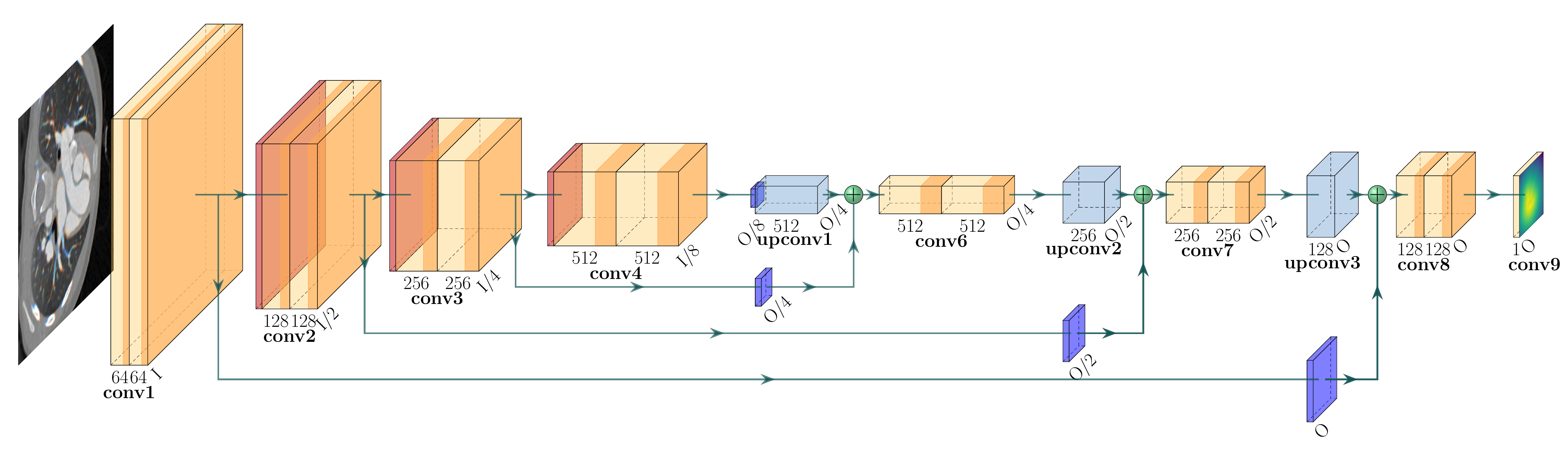}
    \caption{The VGG-Implicit$_1$ (top left), VGG-Implicit$_2$ (top right) and UNetImplicit with depth $d=3$ (bottom) network architectures, mapping CT slices to a spline coefficient grid. The yellow blocks are convolutional layers, with $3\times 3$ kernel and stride~1 followed by batch normalization and a ReLU (ochre), or with $1\times 1$ kernel and $\tanh$ (green) or linear activation, the latter in the final block.
    The lightblue blocks are up-convolutions. Each (up-)convolutional block is labelled with its number of filters (upright) and output spatial resolution (slanted). Each orange block is a $2\times 2$ max pooling, while the darkblue blocks are adaptive average poolings.}
    \label{fig:architectures}
\end{figure*}

\subsection{Network architecture}\label{sec:networkarchitecture}
We have experimented with three different neural network architectures, adapted from existing models. These are shown in Figure \ref{fig:architectures}. We summarize the networks as follows:
\begin{itemize}
    \item VGG-Implicit$_1$: This network is a simple truncation of the VGG-16 network from \cite{simonyan2014very} after the third convolutional block. Two new convolutional layers are then added to reduce the number of channels first from 256 to 16, and then from 16 to one, the first of which has batch normalization and $\tanh$ activations. These added convolutional layers use a kernel of size $1\times1$, introducing a non-linear combination of the features in each location to predict the value of the corresponding spline control point.
    Given that this network is heavily truncated, the extra non-linear activation is used to slightly increase the expressiveness of the network.
    VGG-Implicit$_1$ is a very compact network, which can be used for fast evaluation. The output resolution is constrained to be one quarter of the size of the input in each direction, so $512\times512$ input gives an output grid of $128\times128$ coefficients.
    \item VGG-Implicit$_2$: This network is similarly truncated, but this time after the fourth convolutional block. Again a convolutional layer with batch normalization and $\tanh$ activation is added to reduce the number of channels from 512 to 32, followed by a final convolutional layer that reduces the number of channels to 1. Both these layers use $1\times1$ kernels. This provides a deeper network with larger receptive field. The output resolution of VGG-Implicit$_2$ is constrained to be one eighth of the size of the input in each direction, so $512\times512$ input gives an output grid of $64\times64$ coefficients.
    \item UNetImplicit: To further increase the receptive field and depth of the network, we also adapt the UNet architecture from \cite{ronneberger2015u} to our setting. Since, in principle, we are interested in arbitrary resolution input (images) but fixed resolution output (spline coefficients), we have implemented changes that make the network adaptive. This is done through several applications of adaptive average pooling layers. 
    
    In the original UNet paper, `copy and crop' skip connections are used to attach the outputs of the convolutional blocks on the contracting path to the corresponding blocks on the expansive path. In UNetImplicit, we use adaptive average pooling instead of cropping to allow for freedom of choice in the output resolution. We also add an adaptive average pooling layer (of size $b\times b$) at the bottleneck of the UNet, before the first up-convolution. 
    Each up-convolution on the expansive path then doubles the resolutions until we reach the desired output resolution. 
    In order to adapt this network to different output resolutions, we can vary both the bottleneck size $b$ and the depth $d$, which we define as the number of down-scalings (max-poolings), which we always keep the same as the number of up-scalings (up-convolutions).
    For example, setting $b=8$ and $d=4$, an input of $512\times512$ will produce an output of $128\times128$.
    To reduce the output to $64\times64$, we can either set $b=4$ or set $d=3$. 
    Another change from the original UNet is that padding is used throughout to ensure consistent down- and up-scaling of the input tensor. 
    We also experiment with reducing the number of filters in each convolutional layer, which vastly reduces the total number of training parameters in the network.

\end{itemize}
In addition to these networks, we experimented with bottleneck architectures of encoder-decoder type, but these yielded sub-optimal results, likely due to a lack of spatial awareness.
Further details of the network implementation can be found in the code repository \cite{Barrowclough2020}.

\subsection{Loss functions}\label{ssec:loss}
Because the networks we use are designed to output a specific spline coefficient resolution $O$ (in each direction) that is smaller than the ground truth mask resolution $I = 512$, the first step in each loss function is to evaluate the spline specified by the predicted coefficients at uniformly spaced parameters. This makes it possible to compare arrays of the same shape. The general approach to spline evaluation as presented in Section \ref{ssec:splinemodelling} can be implemented as a recursive algorithm, known as the Cox-De Boor algorithm. However, spline evaluation can also be reduced to a simple matrix multiplication, which is much easier to implement in a way that is compatible with the automatic differentiation used to compute the gradient during back-propagation. This is particularly simple for open uniform knot vectors and repeated evaluation of the same points. To implement this evaluation, we first precompute a single \emph{univariate collocation matrix} (assuming both the spline coefficient arrays and the masks have identical length in each direction): 
\[
U := \big(B_{k}(s_i)\big)_{i=1,k=1}^{I,O},\qquad s_i := \frac{(O-p)(i-1)}{I-1},\qquad i = 1,\ldots,I.
\] 
This collocation matrix is computed on the CPU a single time before training, after which it is passed to the GPU for all future evaluations.

\begin{remark}
For certain special combinations of the input resolution $I$, output resolution $O$ and degree $p$, the collocation matrix has a repeating structure. In these cases, the spline evaluation becomes a special case of an up-convolution in which the weights are cardinal B-splines evaluated at uniformly spaced intervals. We have chosen to use a precomputed collocation matrix rather than adding an up-convolutional layer with fixed weights for the reasons described in Section \ref{sec:geometricrepresentations}, and to allow for flexible combinations of $I, O$, and $p$.
\end{remark}

The actual application of the collocation matrix $U$ to evaluation of a batch of $B$ bivariate spline coefficient arrays, represented as a triple array $C \in \RR^{B, O, O}$,  is done using a two-stage application of Einstein summation, which is available in PyTorch.
The first application creates an intermediate tensor $\tilde{Z} \in \RR^{B,I,O}$ as
\[
\tilde{Z}_{mil} = \sum_k U_{ik} C_{mkl},
\]
followed by creation of the final tensor $Z\in \RR^{B,I,I}$ as
\[
Z_{mij} = \sum_l U_{jl} \tilde{Z}_{mil}.
\]
Here $m$ indexes a single element in the current batch of predicted coefficient arrays.
Note that applying the tensor contraction in two stages via use of a univariate collocation matrix is necessary to ensure both memory and computational efficiency.

Once the predicted tensor $C$ has been converted to a new tensor $Z$ that matches the dimensions of the batch of ground truth masks, we can apply a number of different loss functions.

The simplest approach is to apply traditional loss functions such as mean average error (MAE) loss or mean squared error (MSE) loss. Since we are interested in the zero set of the spline function, it is natural to first transform the mask to contain values in $\{-1,1\}$ before applying MAE or MSE loss. We thus name our loss functions \emph{mask-MAE} (MMAE) and \emph{mask-MSE} (MMSE), in order to distinguish from a direct MAE/MSE loss on the coefficients. 
After mapping the ground truth mask $Y \in \{0,1\}^{I, I}$ to $\hat{Y} := 2Y-1\in\{-1, 1\}^{I, I}$, these loss functions, with mean reduction over the entire tensor, can be given as
\[
L_{\text{MMAE}} := \text{mean}\big(|Z - \hat{Y}|\big),
\]
and
\[
L_{\text{MMSE}} := \text{mean}\big((Z - \hat{Y})^2\big),
\]
respectively. Here, all tensor operations are applied elementwise.

Alternatively, we can base the loss functions on relevant metrics that are used for image segmentation, such as the Jaccard index, the Dice similarity coefficient, and the accuracy.
The predicted binary segmentation is readily deduced from the predicted implicit form as the sign of the evaluated implicit form in each pixel. Hence, the resulting segmentations can be evaluated on a pixel-by-pixel basis; in each pixel, the prediction can be classified into four different categories as true positive (\textit{TP}), true negative (\textit{TN}), false positive (\textit{FP}) and false negative (\textit{FN}).
The \emph{Jaccard index} (Jaccard), also known as the \emph{intersection over union}, is defined as the intersection between the predicted image and the manual reference segmentation divided by their union, that is:
\begin{equation}
\label{eq:jacc}
\text{Jaccard}  :=  \dfrac{TP}{TP + FP + FN}.
\end{equation}
The \emph{Dice similarity coefficient} (Dice) is a measure of the spatial overlap between the predicted image and the manual reference segmentation, written as:
\begin{equation}
\label{eq:dice}
\text{Dice}   :=  \dfrac{2TP}{2TP + FP + FN}.
\end{equation}
The \emph{accuracy} (Accuracy) is a measure of the closeness between the predicted image and the manual reference segmentation, written as:
\begin{equation}
\text{Accuracy}   :=  \dfrac{TP + TN}{TP + FP + FN + TN}.
\end{equation}

Based on these metrics, we define three new loss functions $L_{\text{Jaccard}}$, $L_{\text{Dice}}$, and $L_{\text{Accuracy}}$. These definitions are compatible with the automatic differentiation used during back-propagation. The first stage in computing these losses is to first transform the tensor $Z$ into a tensor $\hat{Z}$ only containing zeros and ones:
\[
\hat{Z} := \frac12 \left(\frac{Z}{\varepsilon+|Z|} + 1\right) ,
\]
where $\varepsilon$ is a small number (typically 0.0001) used to avoid numerical issues with potential division by zero.
We can now define the losses as:
\begin{align}
L_{\text{Jaccard}} & := 1 - \frac{\text{sum}(Y\hat{Z})}{\text{sum}(Y+\hat{Z}-Y\hat{Z})},\notag\\
L_{\text{Dice}} & := 1 - \frac{\text{sum}(2Y\hat{Z})}{\text{sum}(Y+\hat{Z})}, \label{eq:diceloss}\\
L_{\text{Accuracy}} & := 1 - \frac{\text{sum} (\textbf{1} - Y - \hat{Z} + 2Y\hat{Z})}{\text{sum}(\textbf{1})}, \notag
\end{align}
where sums are taken over the entire tensor and $\textbf{1}$ denotes the tensor with dimensions identical to $Y$ and $\hat{Z}$ and all entries equal to 1.

\subsection{Training}
\label{ssec:train}
For training the model, we have used an Intel(R) Core (TM) i7-7700K CPU 4.20GHz (8 cores), 64GB RAM, and NVIDIA Geforce GTX 1080 Ti - PCIE - 11GB of VRAM. We chose a batch size of 10, in order to have a consistent batch size between experiments. The limiting factor for the batch size was the GPU memory available for the higher resolution inputs and outputs. The networks were trained with a stochastic gradient descent optimizer with Nesterov momentum \cite{Sutskever2013} of 0.9 and a learning rate of 0.001. These values were set after some early experimentation to determine parameters that generally worked well. The loss functions used were as described in the previous section.

\section{Experiments and results}
We evaluate our approach under several different metrics in order to compare and determine optimal parameters for this dataset. The data parameters we consider are bidegree $(p, p)$ and output resolution $O$, as well as parameters that determine the architecture of the UNetImplicit network (bottleneck size $b=4,8$, number of spatial down- and up-scalings $d=3,4,5$ and number of filters per convolutional layer). We also compare our approach with state-of-the-art results.

\begin{figure}
\centering
\noindent\begin{tabular}{rccccc}
 \multicolumn{3}{c}{\qquad\quad Image \qquad\qquad Mask \qquad\qquad   Prediction \hfill\ } &  \multicolumn{3}{c}{\qquad\quad Image \qquad\qquad Mask \qquad\qquad   Prediction \hfill\ } \\
\multicolumn{3}{c}{ \includegraphics[scale=0.14]{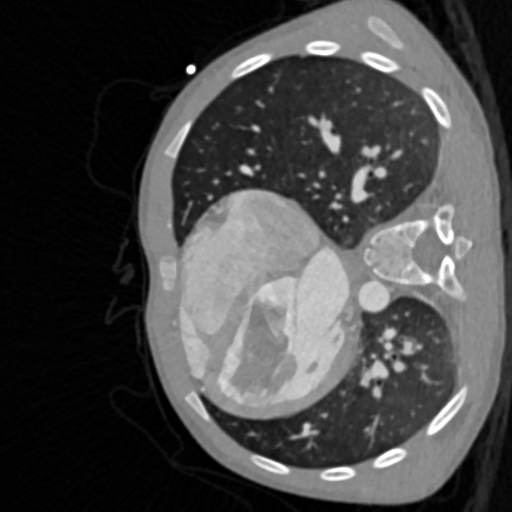}  \includegraphics[scale=0.14]{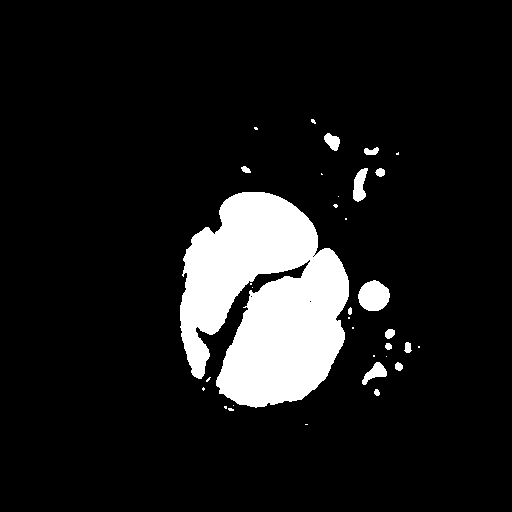} \includegraphics[scale=0.14]{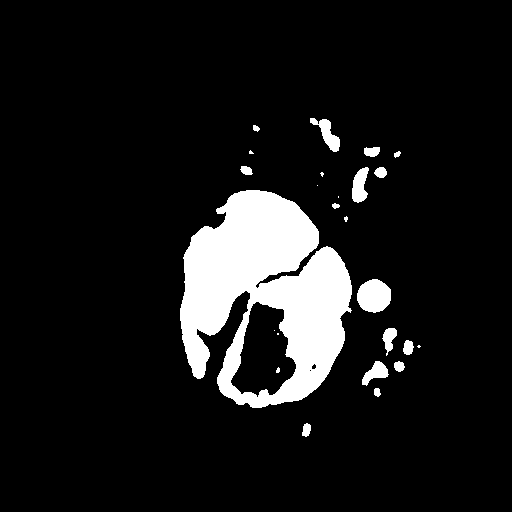}} & 
\multicolumn{3}{c}{
 \includegraphics[scale=0.14]{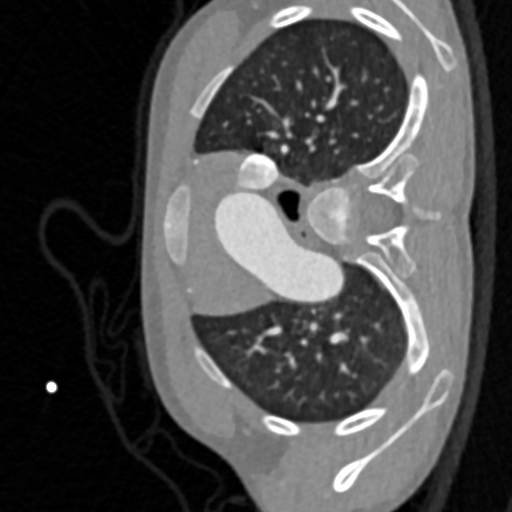}  \includegraphics[scale=0.14]{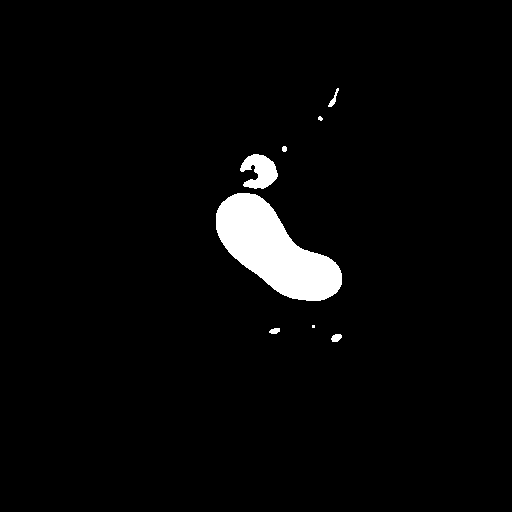} \includegraphics[scale=0.14]{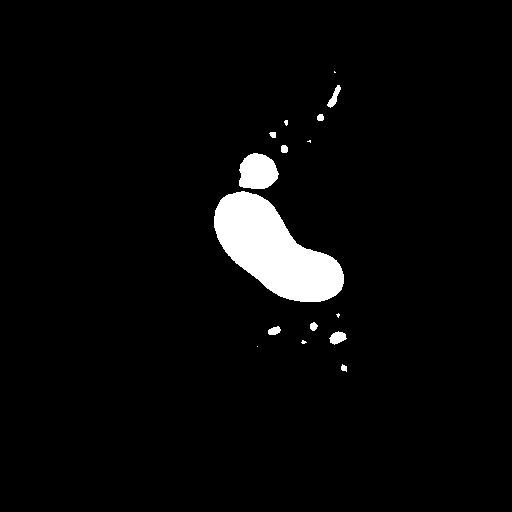}} \\
 \multicolumn{3}{c}{
 \includegraphics[scale=0.14]{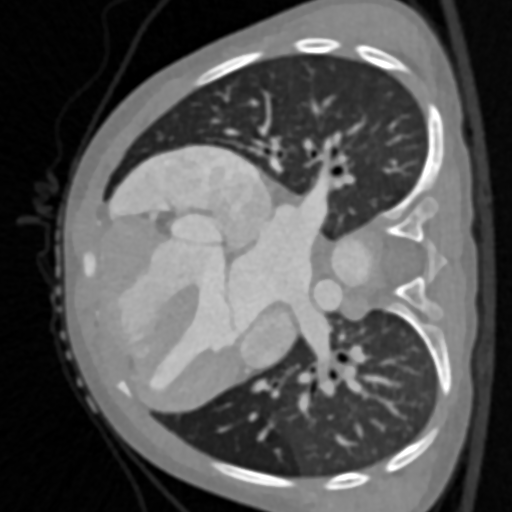}  \includegraphics[scale=0.14]{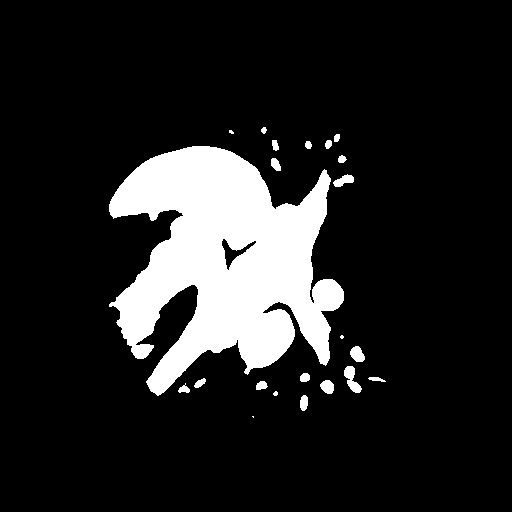}  \includegraphics[scale=0.14]{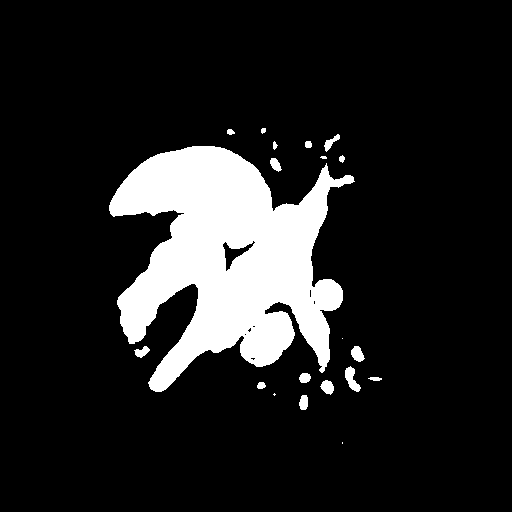}} & 
 \multicolumn{3}{c}{
 \includegraphics[scale=0.14]{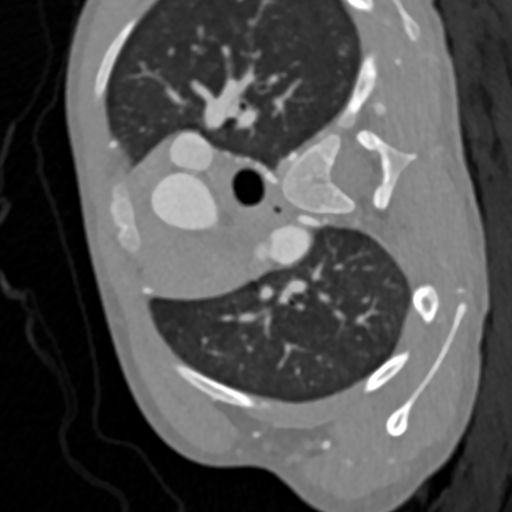}  \includegraphics[scale=0.14]{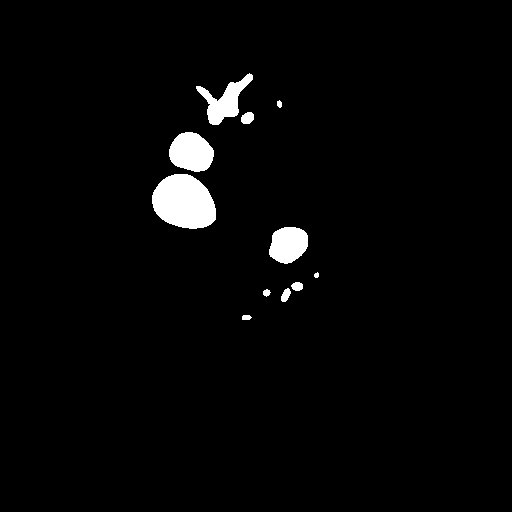}  \includegraphics[scale=0.14]{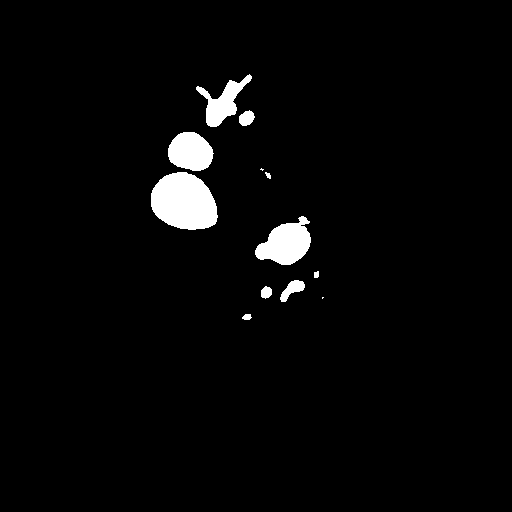}} \\
\multicolumn{3}{c}{
 \includegraphics[scale=0.14]{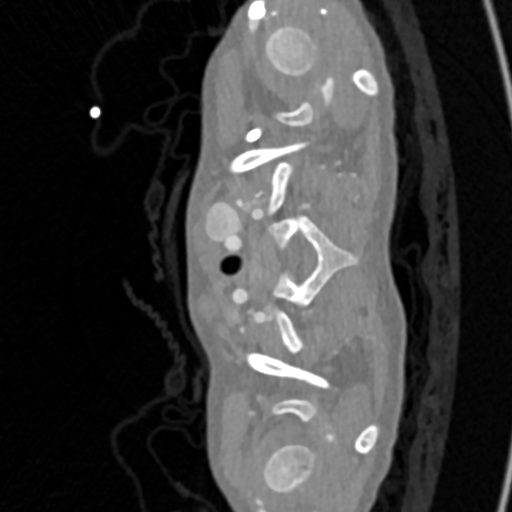}  \includegraphics[scale=0.14]{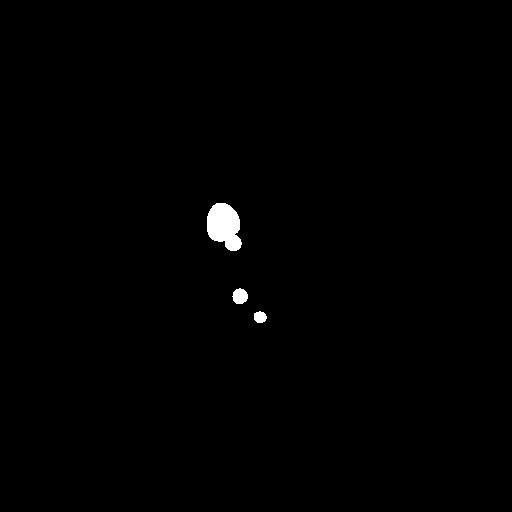}  \includegraphics[scale=0.14]{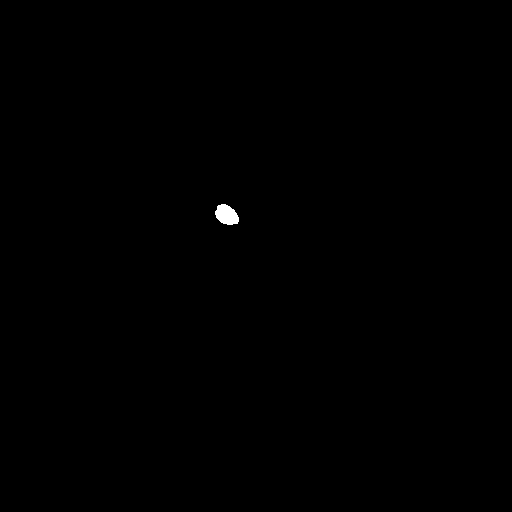}} &
\multicolumn{3}{c}{ 
 \includegraphics[scale=0.14]{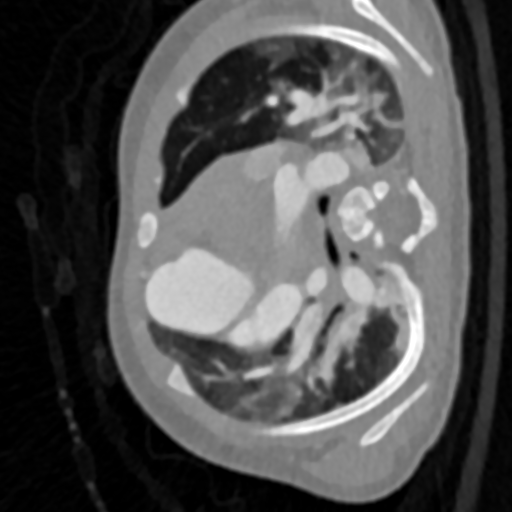}  \includegraphics[scale=0.14]{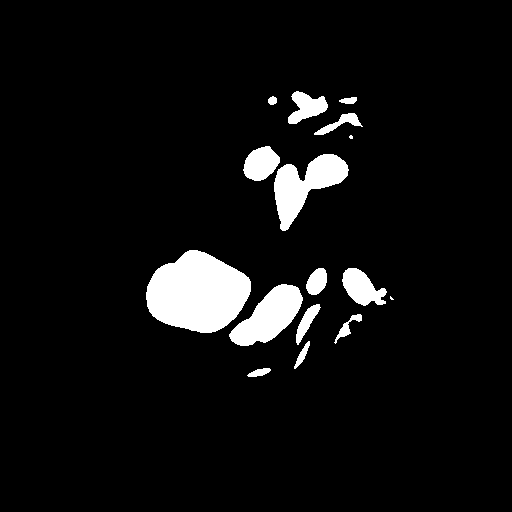}  \includegraphics[scale=0.14]{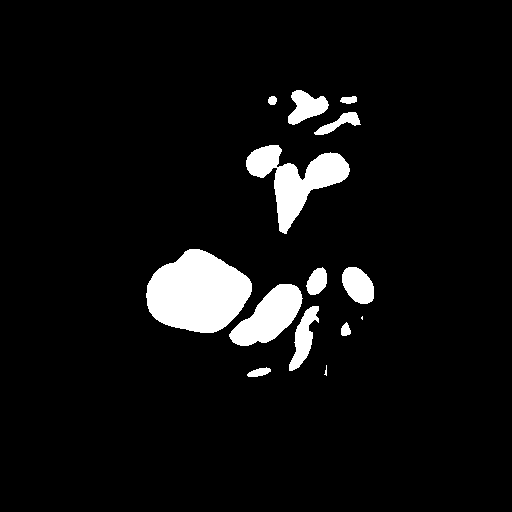}}
\end{tabular}
\caption{Randomly selected images that show the variability of the test dataset, together with manually labelled masks and predictions using UNetImplicit with depth $d=4$, bottleneck size $b=8$, and degree $p=1$, trained for 100 epochs.}\label{fig:slices}
\end{figure}

\subsection{Evaluation metrics}
We now turn our attention to the metrics used for evaluating our networks. Above we have described the accuracy, Dice index and Jaccard index. Contrary to the Dice and Jaccard scores, the accuracy is symmetric in true positives and true negatives. However, the dataset we consider contains large regions of negatives, meaning that in many cases high accuracy scores can be achieved just by predicting blank masks. Thus the high accuracies achieved should only be considered relative to other methods implemented on this dataset. 

Next we consider another evaluation metric not mentioned in Section \ref{ssec:loss}, as its computation is too slow for effective use as a loss function. For any subsets $\YYY, \ZZZ$ of a metric space with metric $d$, one defines the \emph{Hausdorff distance} (HD) as
\begin{align}
\text{HD}(\YYY,\ZZZ) = \max\left[ \adjustlimits \sup_{y \in \YYY} \inf_{z \in \ZZZ} d(x,y),\; \adjustlimits \sup_{z \in \ZZZ} \inf_{y \in \YYY} d(x,y)  \right],   
\end{align} 
where $\sup$ and $\inf$ represent the supremum and infimum, respectively; see \cite{hs}. In our experiments, for the Euclidean distance $d$ of $\RR^3$, we compute the (3D) Hausdorff distance $\text{HD}(\mathcal{Y},\mathcal{Z})$ of
\[\mathcal{Y} := \{(i,j,\ell): Y^\ell_{ij} = 1\}\]
and 
\[\mathcal{Z} := \{(i,j,\ell): F^\ell(i,j) < 0\},\]
where $Y^\ell$ is the mask of the $\ell$-th layer in a single volume and $F^\ell$ is the tensor-product spline function defined by the predicted spline coefficients $C^\ell$ on that layer.
Smaller values of Hausdorff distance correspond to better segmentation accuracy. However, it should be noted that even small regions of false positives can cause large Hausdorff distances, if the false positive is far from the ground truth in pixel space.

\subsection{Model selection}
We have performed a hyperparameter study to identify the contribution of the individual network hyperparameters to the performance of the network, as well as their optimal values. The hyperparameters considered were the spline degree $p$, network depth $d$, bottleneck size $b$ (or equivalently output resolution $O = b\cdot 2^d$), as well as the number of filters in the first layer (which we use to determine the number of filters in subsequent layers).

We observed that in almost all training runs, the validation curve has flattened out after 100 epochs. In addition, the validation curve did not significantly increase during any of our training runs, indicating that any overfitting is minimal. We thus performed most of our experiments by training the networks for 100 epochs, which takes approximately 11--15 hours on our hardware, depending on the network configuration.

\begin{table*}
    \centering
\noindent\begin{tabular*}{\textwidth}{@{\extracolsep{\stretch{1}}}*{8}{c}}
\toprule
Output res. & Depth & \#Parameters &  Size (Mb) & Degree & Val. Accuracy & Val. Dice & Val. Jaccard \\
\midrule
$O =  64$ & $d = 3$ &   7,701,825 &  29.44 & $p = 0$ & 0.9854 & 0.8659 & 0.7647 \\
          &         &             &        & $p = 1$ & 0.9918 & 0.9250 & 0.8610 \\
          &         &             &        & $p = 2$ & 0.9908 & 0.9195 & 0.8515 \\
\cmidrule{2-8}
          & $d = 4$ &  31,042,369 & 118.51 & $p = 0$ & 0.9860 & 0.8741 & 0.7778 \\
          &         &             &        & $p = 1$ & 0.9916 & 0.9238 & 0.8591 \\
          &         &             &        & $p = 2$ & 0.9914 & 0.9242 & 0.8597 \\
\midrule
$O = 128$ & $d = 4$ &  31,042,369 & 118.51 & $p = 0$ & 0.9907 & 0.9162 & 0.8460 \\
          &         &             &        & $p = 1$ & \textbf{0.9926} & \textbf{0.9332} & \textbf{0.8754} \\
          &         &             &        & $p = 2$ & 0.9920 & 0.9310 & 0.8715 \\
\cmidrule{2-8}
          & $d = 5$ & 124,385,089 & 474.64 & $p = 0$ & 0.9898 & 0.9128 & 0.8404 \\
          &         &             &        & $p = 1$ & 0.9925 & 0.9320 & 0.8733 \\
          &         &             &        & $p = 2$ & 0.9923 & 0.9311 & 0.8718 \\
\bottomrule
\end{tabular*}
    \caption{A hyperparameter study of the performance (highest emphasized) for various UNetImplicit network architectures trained with $L_{\text{Dice}}$ loss \eqref{eq:diceloss} for 100 epochs, for input resolution $I=512$, output coefficient resolutions $O=64,128$, depth $d=3,4,5$, and spline degrees $p=0,1,2$. All scores are for batches of validation data.} 
    \label{tab:ablation-dpO}
\end{table*}

\begin{figure}[h!]
\centering
\subfigure[]{\includegraphics[scale=0.28]{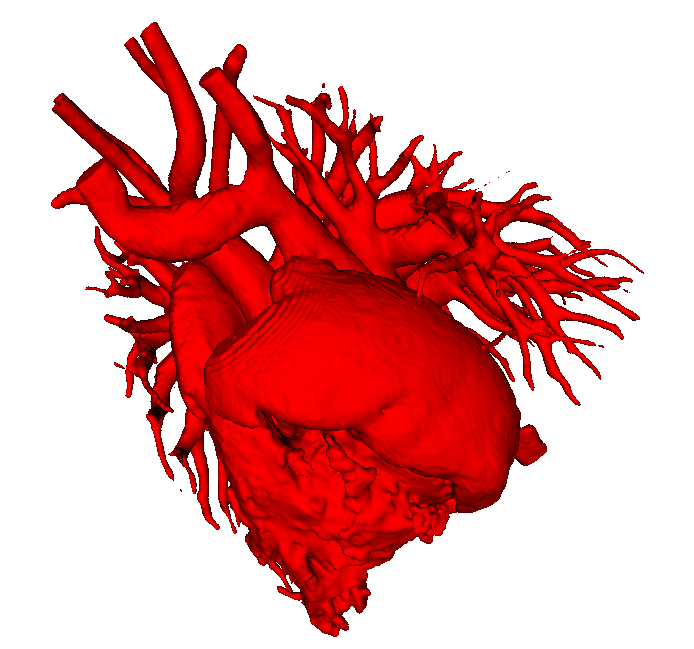} }
\subfigure[]{\includegraphics[scale=0.28]{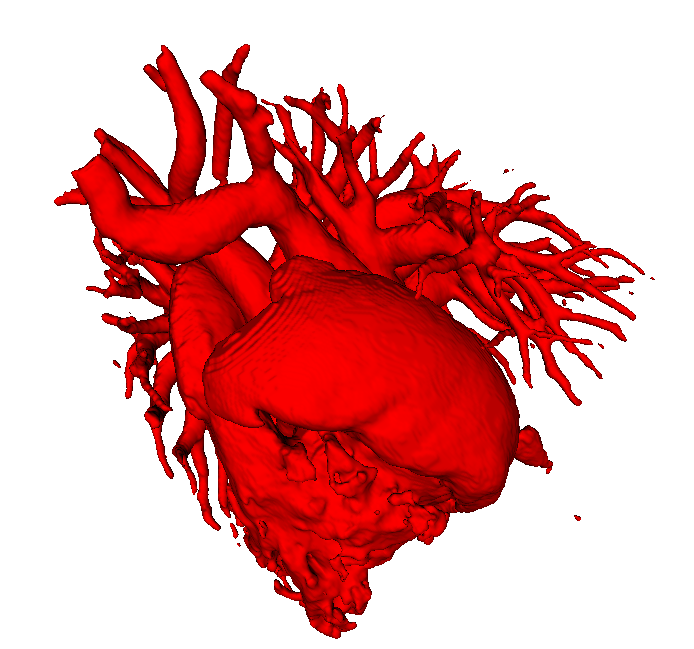} }
\caption{Comparison of (a) manual segmentation and (b) UNetImplicit network prediction for Volume 18, which has highest Dice score in the test dataset} \label{fig:best3D}
\end{figure} 
 \begin{figure}[h!] 
\centering
\subfigure[]{\includegraphics[scale=0.3]{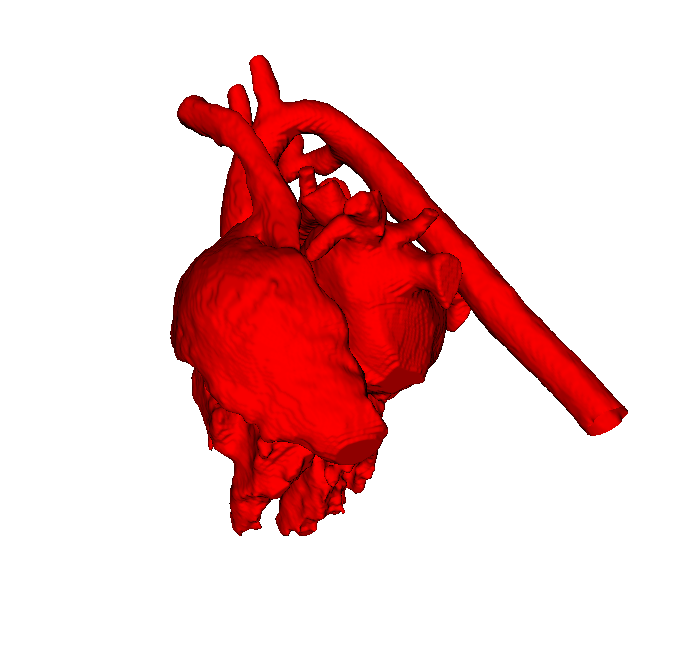} \label{fig:vol40mask}}
\subfigure[]{\includegraphics[scale=0.3]{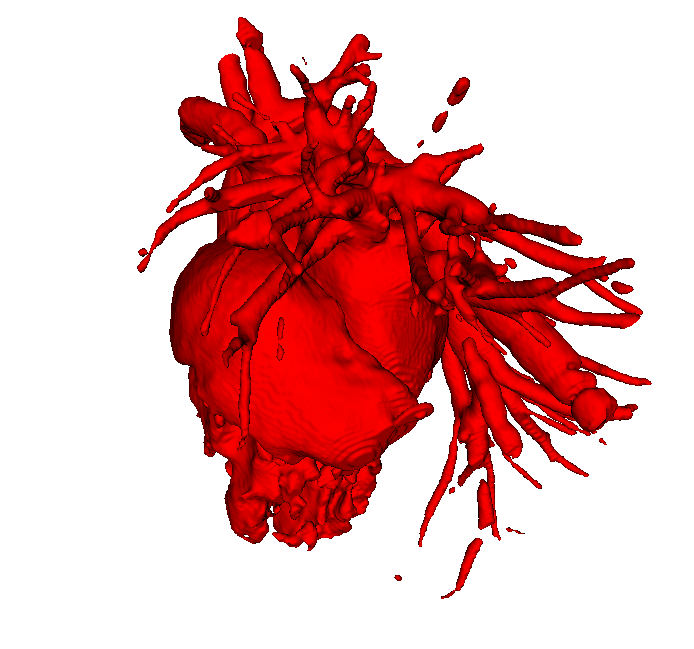} \label{fig:vol40prediction} }
\caption{Comparison of (a) manual segmentation and (b) UNetImplicit network prediction for Volume 40, which has lowest Dice score in the test dataset}  \label{fig:bad3D}
\end{figure} 

Our early experiments showed that UNetImplicit generally provided better results than the VGG-inspired networks. 
The UNetImplicit network can be defined with different depth $d$ and bottleneck size $b,$ and we performed a study to determine the optimal parameters. In Table \ref{tab:ablation-dpO}, we summarize our experiments on the performance of UNetImplicit under these different parameters. Note that varying the depths and bottleneck size changes the coefficient output resolution. In this table, we also consider the performance over different B-spline bidegrees $p=0,1,2$.

Note that the scores presented in this table are based on the validation dataset and correspond to scores computed over a batch of input layers. Hence they are not directly comparable with the scores presented in Tables \ref{tab:evaluation-by-volume} and \ref{tab:model_evaluation}, which are taken over entire volumes. The timings presented in the table are averages and standard deviations of per-layer inference runtimes, when performed with a batch size of one. To compute these timings we utilized the same hardware as described in Section \ref{ssec:train}.

The results of Table \ref{tab:ablation-dpO} show that the UNetImplicit network with $d=4$, $b=8$, and $p=(1,1)$ performs better than the other networks. 
This network also performs better than both the VGG-inspired networks when tested with corresponding parameters, see Table \ref{tab:model_evaluation}. 
We have thus selected UNetImplicit with these parameters as the best model for this dataset. Figure~\ref{fig:slices} shows the predictions made by this model for randomly selected images.

\begin{figure}
    \centering
\subfigure[Vol. accuracy]{
\begin{tikzpicture}[scale=0.75]
  \begin{axis}
    [
    ytick={1,2,3},
    x tick label style={
    /pgf/number format/.cd,
    fixed,
    fixed zerofill,
    precision=3
    },
    yticklabels={VGG-Implicit$_1$, VGG-Implicit$_2$,  UNetImplicit},
    ]
    \addplot+[
    boxplot prepared={
      median=0.9891,
      upper quartile=0.9934,
      lower quartile=0.9871,
      upper whisker=0.9951,
      lower whisker=0.9852
    },
    ] coordinates {};
    \addplot+[
    boxplot prepared={
      median=0.9908,
      upper quartile=0.9929,
      lower quartile=0.9883,
      upper whisker=0.9947,
      lower whisker=0.9855
    },
    ] coordinates {};
    \addplot+[
    boxplot prepared={
      median=0.9927,
      upper quartile=0.9951,
      lower quartile=0.9902,
      upper whisker=0.9971,
      lower whisker=0.9836
    },
    ] coordinates {};
  \end{axis}
\end{tikzpicture}
}
\subfigure[Vol. Dice]{
\begin{tikzpicture}[scale=0.75]
  \begin{axis}
    [
    ytick={1,2,3},
    yticklabels={ , , },
    ]
    \addplot+[
    boxplot prepared={
      median=0.9026,
      upper quartile=0.9218,
      lower quartile=0.8722,
      upper whisker=0.9334,
      lower whisker=0.8380
    },
    ] coordinates {};
    \addplot+[
    boxplot prepared={
      median=0.9097,
      upper quartile=0.9201,
      lower quartile=0.8918,
      upper whisker=0.9289,
      lower whisker=0.8123
    },
    ] coordinates {};
    \addplot+[
    boxplot prepared={
      median=0.9252,
      upper quartile=0.9398,
      lower quartile=0.8901,
      upper whisker=0.9611,
      lower whisker=0.8167
    },
    ] coordinates {};
  \end{axis}
\end{tikzpicture}
}
\subfigure[Vol. Jaccard]{
\begin{tikzpicture}[scale=0.75]
  \begin{axis}
    [
    ytick={1,2,3},
    yticklabels={, , },
    ]
    \addplot+[
    boxplot prepared={
      median=0.8225,
      upper quartile=0.8549,
      lower quartile=0.7735,
      upper whisker=0.8751,
      lower whisker=0.7211
    },
    ] coordinates {};
    \addplot+[
    boxplot prepared={
      median=0.8344,
      upper quartile=0.8520,
      lower quartile=0.8047,
      upper whisker=0.8672,
      lower whisker=0.6839
    },
    ] coordinates {};
    \addplot+[
    boxplot prepared={
      median=0.8609,
      upper quartile=0.8864,
      lower quartile=0.8020,
      upper whisker=0.9251,
      lower whisker=0.6902
    },
    ] coordinates {};
  \end{axis}
\end{tikzpicture}
}
\subfigure[Vol. Hausdorff]{
\begin{tikzpicture}[scale=0.75]
  \begin{axis}
    [
    ytick={1,2,3},
    yticklabels={, , },
    ]
    \addplot+[
    boxplot prepared={
      median=139.4041,
      upper quartile=152.6666,
      lower quartile=119.7245,
      upper whisker=214.7696,
      lower whisker=68.7750
    },
    ] coordinates {};
    \addplot+[
    boxplot prepared={ 
      median=126.7757,
      upper quartile=133.9604,
      lower quartile=116.6745,
      upper whisker=228.4601,
      lower whisker=80.3057
    },
    ] coordinates {};
    \addplot+[
    boxplot prepared={
      median=96.4317,
      upper quartile=123.6252,
      lower quartile=81.0493,
      upper whisker=212.7581,
      lower whisker=34.7275
    },
    ] coordinates {};
  \end{axis}
\end{tikzpicture}
}
    \caption{For various networks and metrics, box plots of the values of these metrics for the volumes in the test data set.}
    \label{fig:boxplots}
\end{figure}
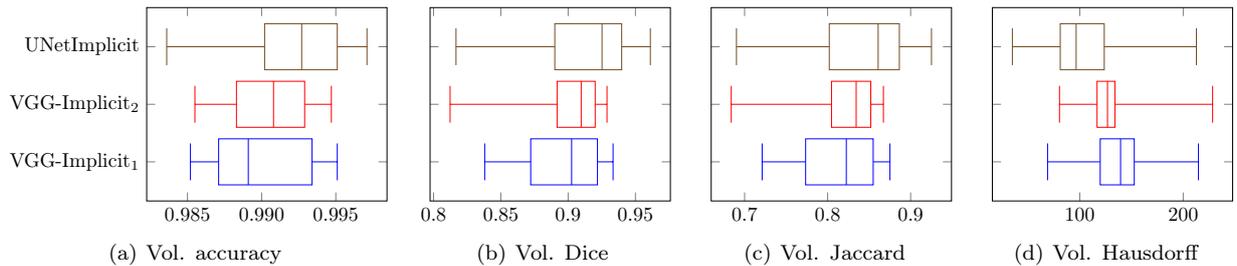

The box plots in Figure \ref{fig:boxplots} show the distribution of scores for the different networks, including the minimum, lower quantile, median, upper quantile and maximum scores over the test volumes. They confirm that, in general, UNetImplicit does indeed perform significantly better than the VGG-inspired networks. However, it should be noted that the lowest accuracy and the second-lowest Dice and Jaccard scores were obtained using UNetImplicit. This suggests that in some edge cases the VGG networks can perform better. It is interesting to note that VGG-Implicit$_2$ on average performs slightly better than VGG-Implicit$_1$, despite it having output resolution $O=64$, which is half that of VGG-Implicit$_1$. This could be explained by the fact that VGG-Implicit$_2$ includes one convolutional block more than VGG-Implicit$_1$, resulting in a deeper network with larger receptive field.

After training for 320 epochs, we managed to improve the results for UNetImplicit with $d=4,$ $b=8,$ even further (to an average Dice score of $0.936$ on batches of the validation data).
The results of this network for each individual test volume is shown in Table \ref{tab:evaluation-by-volume}. 
For these results, the predicted 2D images corresponding to each CT volume are combined to make full 3D volumes.
Figures \ref{fig:best3D} and \ref{fig:bad3D} show 3D views of the volumes with best Dice score (0.9656) and worst Dice score (0.8222), respectively. 

The lowest Dice score and highest Hausdorff distance were observed for Volume 40 in the test dataset. However, in this case, the poor scores can be explained by the fact that the manual segmentation has not included all branches of the pulmonary artery, as shown in Figure \ref{fig:vol40mask}. 
Since our model is trained with CT images that include the blood vessels of the pulmonary artery, these are picked up in the predictions, see Figure \ref{fig:vol40prediction}. 
Hence, the network exhibits over-segmentation, and the Dice score and Hausdorff distance suffer accordingly. 
We also observed some disconnected components in the predicted 3D model. 
Since the network is trained and tested on 2D images, some of the slices with few positives exhibit more variability in the quality of the predictions, which may be the main cause of these disconnected components.

We performed some experiments to examine the effect of reducing the number of filters. For the sake of brevity we avoid presenting detailed results here, but we observed that halving the number of filters at each convolution led to a modest reduction in the overall score. Reducing to one quarter of the original number of filters reduced the score further, but the results could still be considered reasonable. Given that reducing the number of filters vastly reduces the size of the networks, reducing the number of filters at a minor expense of accuracy could be considered beneficial in some applications (e.g. real-time segmentation for dynamic visualization).

\begin{remark}
Note that the results in this section are before applying many of the standard ``tricks'' for boosting performance. In particular, further improvements might be possible by applying appropriate data augmentation and ensemble modelling. Removing small components that are disconnected from the main structures may also improve the scores, especially with respect to Hausdorff distance.
\end{remark}

\begin{table}
    \centering
\noindent\begin{tabular*}{\textwidth}{l@{\extracolsep{\stretch{1}}}*{4}{c}}
\toprule
Test volume number  & Vol. Accuracy & Vol. Dice   &  Vol. Jaccard   &  Vol. Hausdorff \\
\toprule
Volume 02 & 0.9961 & 0.9516 & 0.9076 & \ 48.6 \\
Volume 12 & 0.9852 & 0.8866 & 0.7963 & 104.4 \\
Volume 18 & \textbf{0.9975} & \textbf{0.9656} & \textbf{0.9335} & \textbf{\ 25.8} \\
Volume 23 & 0.9915 & 0.9412 & 0.8890 & 104.0 \\
Volume 25 & 0.9955 & 0.9250 & 0.8605 & \ 53.8 \\
Volume 27 & 0.9937 & 0.9341 & 0.8763 & \ 74.1 \\
Volume 40 & 0.9850 & 0.8222 & 0.6980 & 199.8 \\
Volume 44 & 0.9903 & 0.9331 & 0.8747 & 173.6 \\
Volume 48 & 0.9958 & 0.9024 & 0.8221 & \ 74.4 \\
Volume 51 & 0.9919 & 0.9271 & 0.8641 & \ 58.7 \\
Volume 53 & 0.9912 & 0.9068 & 0.8295 & \ 87.8 \\
Volume 55 & 0.9956 & 0.9465 & 0.8984 & \ 31.6 \\
Volume 57 & 0.9944 & 0.9414 & 0.8894 & \ 94.2 \\
Volume 60 & 0.9846 & 0.8657 & 0.7633 & 145.1 \\
\midrule
Average & 0.9920 & 0.9178 & 0.8502 & \ 91.1 \\
\midrule
Standard deviation & 0.0042 & 0.0370 & 0.0609 & \ 49.7 \\
\bottomrule
\end{tabular*}
    \caption{3D (volumetric) Accuracy, Dice, Jaccard and Hausdorff scores for UNetImplicit (trained for 320 epochs) applied to each of the test volumes, considered individually.}
    \label{tab:evaluation-by-volume}
\end{table}

\subsection{Loss functions}

We trained the UNetImplicit network with depth $d=4$, bottleneck size $b=8$, and degree $p=1$ with the different loss functions described in Section \ref{ssec:loss}. It is hard to compare the obtained validation losses directly, as validation loss will typically favour the loss function that it is trained on. However, besides a qualitative comparison, it is possible to train on each loss and then evaluate on all the other loss functions, as shown in Table \ref{tab:ablationloss}. This approach is inspired by the principle of cross-validation.

Table \ref{tab:ablationloss} shows that training with accuracy, Dice and Jaccard loss all achieve high scores on each other's evaluation metrics. Likely due to this, training on linear combinations of Dice and Jaccard loss did not improve the scores.
However, training with accuracy, Dice and Jaccard loss does not give good scores with respect to the MMSE and MMAE evaluation metrics. This is expected, as the MMSE and MMAE metrics are tailored to approximating functions with values in $\{-1,1\},$ and this constraint is not imposed on the results when using these loss functions. The MMSE and MMAE losses are also very similar, both achieving high scores in all evaluation metrics, albeit with slightly lower scores on the accuracy, Dice and Jaccard evaluation metrics. When training directly for MSE of the predicted coefficients and precomputed implicit spline approximations of the ground truth segmentation masks, a Jaccard score of around 0.8 was achieved on the validation set.

\begin{table}
    \centering
\noindent\begin{tabular*}{\textwidth}{l@{\extracolsep{\stretch{1}}}*{5}{c}}
\toprule
Training loss & $1 - \text{MMSE}_{\text{val}}$ & $1 - \text{MMAE}_{\text{val}}$ & Accuracy$_{\text{val}}$ & Dice$_{\text{val}}$ & Jaccard$_{\text{val}}$\\
\midrule
$\text{MMSE}$         & $\phantom{+}$0.9744 & $\phantom{+}$0.9685 & 0.9919 & 0.9275 & 0.8654\\
$\text{MMAE}$         & $\phantom{+}$0.9734 & $\phantom{+}$0.9743 & 0.9916 & 0.9234 & 0.8583\\
$1 - \text{Accuracy}$ & $\phantom{+}$0.5506 & $\phantom{+}$0.5701 & 0.9921 & 0.9297 & 0.8691\\
$1 - \text{Dice}$     & $-$3.5035 & $-$0.7471 & 0.9926 & 0.9336 & 0.8761\\
$1 - \text{Jaccard}$  & $-$7.5589 & $-$1.3616 & 0.9927 & 0.9341 & 0.8768\\ \bottomrule
\end{tabular*}
\caption{Performance of UNetImplicit (depth $d=4$, bottleneck size $b=8$, degree $p=1$, trained for 100 epochs) evaluated in various metrics when varying the training loss function.}\label{tab:ablationloss}
\end{table}

\subsection{Comparison to state of the art} 
As far as we are aware, there exist two papers in the literature that make use of the same CHD CT dataset as us. We have organized the performances in terms of various metrics for our network and these networks in Table \ref{tab:model_evaluation}, bearing in mind that it is difficult to make entirely fair comparisons due to the reasons described below. Note that UNetImplicit (with optimal parameters) can process about 200 slices per second, roughly corresponding to a single CT volume per second. Increasing the batch size can be expected to yield a significant speed improvement, in particular for the smaller networks.

\cite{miccai2019} have used a 2D UNet architecture, and observed the average Dice score 0.7843 for blood volume. However, in their approach, blood volume is split into several anatomically separate parts, and the average is taken over the individual Dice scores for these predicted parts. This appears to be a harder problem to solve, and may be the main cause of the lower score observed.

\cite{Rajan} considered a DenseVNet architecture, and obtained the average volumetric Dice score of 0.9183 for blood volume, which is very similar to the results we obtain with UNetImplicit. The other scores for DenseVNet are also presented in Table \ref{tab:model_evaluation}. They achieve a lower average Hausdorff distance than in our approach, and this may in part be explained by the fact that the authors post-process the test results by removing components that are disconnected from the main structure. Since our approach is inherently 2D, we have chosen not to remove components that are disconnected in 3D in this paper.

\begin{table*}
    \centering
\noindent\begin{tabular*}{\textwidth}{l@{\extracolsep{\stretch{1}}}*{5}{c}}
\toprule
Model        & Vol. Accuracy & Vol. Dice & Vol. Jaccard & Vol. Hausdorff & Inference time (ms) \\ \toprule 
VGG-Implicit$_1$ & 0.9902 & 0.8949 & 0.8112 & 139.04 & \textbf{1.14 $\pm$ 0.05} \\
\midrule
VGG-Implicit$_2$ & 0.9904 & 0.8966 & 0.8143 & 130.40 & 1.49 $\pm$ 0.07\\
\midrule
UNetImplicit     & 0.9920 & 0.9178 & 0.8502 & \ 91.10 & 5.56 $\pm$ 2.77 \\
\midrule
DenseVNet        & 0.9959 & 0.9183 & 0.8509  & \ 58.80 & --  \\
\bottomrule
\end{tabular*}
\caption{Model evaluation and comparison of 3D metrics and per-slice average inference time (mean $\pm$ standard deviation) on the blood volume test dataset. The top three networks are trained for 100 epochs, and the parameters used for UNetImplicit are $b=8$, $d=4$, and $p=1$. 
}\label{tab:model_evaluation}
\end{table*}

\section{Conclusion}

\subsection*{Summary of the paper}
In this paper, we have introduced a new method for segmenting image data using implicit spline representations and deep learning. 
We have shown that our approach is effective at segmenting blood volume in a medical imaging dataset and that we are able to achieve state-of-the-art results by using a modified version of the UNet architecture, which we call UNetImplicit. 

By modelling the segmentation boundaries implicitly, we are able to perform segmentation even in the presence of complex topologies, and the use of spline representations ensures a compact representation that can be subsequently sampled at any desired resolution. In the case of our best network, the total number of output spline coefficients is one sixteenth of the total number of input pixels. Besides these representational advantages, our method is also amenable to geometric processing operations. In particular, inside/outside computations are reduced to mere function evaluations, and shapes can be efficiently manipulated and compared in terms of the spline control net.

While we have chosen to focus on medical imaging in this paper, the proposed method is not limited to this application domain. Early experiments performed on the Cityscapes dataset \cite{cordts2016cityscapes} yielded promising results, in which we experienced that a spline grid of $14\times14$ coefficients was sufficient for most segmentations. Hence the resolution of the spline coefficients required to obtain good results depends on the input data, both with respect to variability of the data and the presence of sharp features. This also illustrates that in other application domains a further reduction in representation size is possible.

\subsection*{Limitations}
One of the main limitations of our approach is that we need to used a fixed spline resolution for all samples. This imposes an upper limit on the number of oscillations the spline can have in a given region. Depending on the characteristics of the data, the spline resolution may need to be increased to capture all details. Our approach also has limited ability to model sharp features, in that splines are inherently smooth. However, this is more a theoretical than practical limitation due to uncertainty or blurring at the pixel level, which is typical for imaging datasets (including the CHD CT dataset). Another limitation of using implicit representations is that they can only deal with a single segmentation class per output channel. A potential solution to this is to output multiple segmented classes in separate channels using weight sharing and multitask learning.  

\subsection*{Future work}
We foresee several directions for future work based on our approach. 
In this paper, we have restricted our attention to B-spline basis functions, but the only requirement is that the basis can be evaluated by multiplication with a pre-computed collocation matrix. Thus the method can adopt any basis functions arranged in a grid compatible with adaptive average pooling of the contracting path. This opens up avenues for exploring how different types of basis functions perform on different datasets. 

Thus far we have employed implicit functions solely for the purpose of determining the segmentation boundary. A richer form of implicit function is a signed distance function to the boundary. In that case, the gradient of the implicit function (projected to the plane $z=0$) is the normal direction along the segmentation boundary. Hence different level sets of the implicit (signed distance) function can be used to generate offsets or confidence bounds of the segmentation boundary. While signed distance functions are not typically smooth everywhere, they can be approximated closely by spline functions.

Although we have only considered 2D segmentation in this paper, the approach is directly generalizable to 3D. A 3D implementation will allow us to take advantage of smoothness in the axial direction, which should lead to even more compact representations. In particular, multislice methods could be employed to take advantage of gradients in the slice direction.

Finally, we envisage that our approach can be useful in other application areas. For example, reconstructing digital twins of 3D printed objects from images taken at each layer of the manufacturing process is one potential application. Our approach may also be applied to standard segmentation of photographs if the segmentation boundaries are of generally smooth nature.

\section*{Acknowledgements}
This project was supported by an IKTPLUSS grant, project number 270922, from the Research Council of Norway, as well as by the European Union's Horizon 2020 Research and Innovation Programme under Grant Agreement numbers 951956 and 860843. We would like to thank Dr. Xiaowei Xu from the Department of Computer Science and Engineering, University of Notre Dame, for providing us access to the CHD CT dataset. We wish to thank Arturs Berzins and the anonymous referees for their valuable feedback on an earlier draft of this paper.

\bibliography{chd}

\end{document}